\documentclass{cup-hpl}

\usepackage{graphicx}% Include figure files
\usepackage{dcolumn}% Align table columns on decimal point
\usepackage{bm}% bold math

\usepackage{hyperref}

\usepackage{color}

\usepackage[utf8]{inputenc}
\usepackage[T1]{fontenc}
\usepackage{upgreek}

\begin{document}

\shorttitle{Particle-In-Cell Framework for Modeling Few-Cycle Pulse Damage }
\shortauthor{J.~Smith et al.}

\title{A Fully Three-Dimensional Kinetic Particle-In-Cell Framework for Modeling Laser-Dielectric Interactions: Few-Cycle Pulse Damage }

\author[1]{Joseph R. Smith\corresp{\email{JosephRSmith@protonmail.com}, \email{chowdhury.24@osu.edu}}}  

 \author[2]{Ziyao Su}
\author[2]{Simin Zhang}
\author[3]{Charles Varin}
\author[4]{Vitaly E. Gruzdev}
\author[2]{Enam A. Chowdhury}

\address[1]{Department of Physics, Marietta College, Marietta, Ohio 45750, USA}
\address[2]{Department of Materials Science and Engineering, The Ohio State University, Columbus, Ohio 43210, USA}
\address[3]{C\'egep de l'Outaouais - CyberQuébec, Gatineau, QC, J8Y 6M4, Canada}
\address[4]{Department of Physics and Astronomy, University of New Mexico, Albuquerque, NM, 87106, USA}

\begin{abstract}
We present a fully three-dimensional kinetic framework for modeling intense short pulse lasers interacting with dielectric materials. Our work modifies the open-source Particle-In-Cell (PIC) code EPOCH to include new models for molecular photoionization and dielectric optical response. We use this framework to model the laser-induced damage of dielectric materials by few-cycle laser pulses. The framework is benchmarked against experimental results for bulk silica targets and then applied to model multi-layer dielectric mirrors with a sequence of simulations with varying laser fluence. This allows us to better understand the laser damage process by providing new insight into energy absorption, excited particle dynamics, and nonthermal excited particle distributions.  We compare common damage threshold metrics based on the energy density and excited electron density.

\end{abstract}

\maketitle

\section{Introduction}
The 2023 Nobel Prize in Physics awarded to Pierre Agostini, Ferenc Krausz and Anne L’Huillier ``for experimental methods that generate attosecond pulses of light for the study of electron dynamics in matter'\cite{Nobel2023} and the 2018 National Academies of Science report\cite{national2018opportunities} that spurred the ``Brightest Light Initiative'' \cite{falcone2020workshop} highlight the impactful science enabled by producing the shortest possible pulses of light~\cite{ferray1988multiple,paul2001observation,hentschel2001attosecond}. Recent advances in high-energy few-cycle pulse generation techniques~\cite{Brabec_2000,nagy2021high} allow us to probe new physical effects including those from carrier-envelope phase\cite{wittmann2009singleCEP,PhysRevLett.105.063903_CEP,huijts2021identifyingCEP} and to generate high laser intensities with moderate laser energies. These developments motivate the design of optical components with higher Laser-Induced Damage Thresholds (LIDTs) for few to single-cycle pulses. The scaling of LIDT with laser fluence is generally well understood for the tens of picosecond to nanosecond regimes\cite{Stuart:96}, but is more complex for shorter pulses\cite{Du_1994}, especially few-cycle pulses\cite{Lenzner_1998,Chimier_PhysRevB.84.094104}. 

Understanding few-cycle pulse material interactions require advances in computational and theoretical models. There has been significant work using one-dimensional Finite-Difference-Time-Domain (FDTD) simulations to model these interactions\cite{penano2005transmission}, with recent efforts in two dimensions~\cite{Zhang_2020,zhang2022ultrafast}. FDTD simulations provide insight into the laser material interactions and can predict LIDT, but do not capture the nonthermal nature of excited electrons, nor do they capture their ballistic motion. To expand our understanding of the interaction dynamics, we use Particle-In-Cell (PIC) simulations\cite{BirdsallLangdon2004,hockney1981PIC}. As with FDTD simulations, PIC codes solve Maxwell's equations on a computational grid. Additionally, PIC simulations statistically represent the neutral and excited particles in the simulation with a finite number of `macroparticles' \cite{BirdsallLangdon2004, hockney1981PIC, arber2015contemporary}. The charged particles are then advanced using the Lorentz force and additional physical effects including ionization and collisions are often added. 

PIC simulations are commonly used to study laser-plasma interactions (e.g.,~\citet{ziegler2024laser}) and increasingly being modified to simulate laser damage and related regimes. For example, \citet{Mitchell:15} model crater formation in metals due to femtosecond laser ablation,  \citet{Cochran_2019} model liquid-crystal plasma mirrors \cite{Corkum_1993},  \citet{deziel_2018} study laser-induced periodic surface structures, and  \citet{Ding_plasmon_2020} and \citet{do2021electron} model plasmons using PIC. Interactions of lasers with nano/micro-scale structures in silicon and SiO$_2$ have been modeled with versions of the framework introduced in this work \cite{shcherbakov2023nanoscale,smith2021intense} and recently \citet{charpin2024simulation} developed a similar framework to explore ionization in dielectrics. Our work builds on these efforts specifically for laser damage for few-cycle pulses by including the corrected Keldysh ionization model \cite{keldysh1965ionization,gruzdev_2004,zhang2022ultrafast}, to account for ionization across a range of fluences. We are able explore the nonthermal distribution of excited particles in the laser-damage regime using PIC, which improves our fundamental understanding of damage mechanisms.

In this work, we begin in Sec.~\ref{sec:pic_mod} by introducing the modifications and extensions we have made to a PIC code to model laser-dielectric interactions. Then in Sec.~\ref{sec:setup} we discuss the material properties and simulation setup for both bulk and multi-layer targets. Next, in Sec.~\ref{sec:bulk} we compare the predictions of this framework to existing experimental results and discuss expected damage threshold metrics. Then we apply the framework to the modeling of multi-layer mirrors in Sec.~\ref{sec:multi-layer} and conclude in Sec.~\ref{sec:conclusion}.

\section{Particle-In-Cell Simulation Modifications}\label{sec:pic_mod}
Our work extends the three-dimensional (3D) implementation of version 4.17.10 of the EPOCH~\cite{arber2015contemporary} PIC code, which is designed for the study of high energy density physics. EPOCH is a popular open-source PIC code that can scale to run on thousands of CPU cores, although we note that there are a variety of other open-source and proprietary PIC codes available with different features and implementations\cite{SmithPICCompare2021} such as GPU operation (e.g.,~PIConGPU\cite{PIConGPU} and WarpX\cite{warpx_fedeli_2022}). PIC simulations are typically used to study only tens-to-hundreds of femtosecond timescales due to numerical instability issues and computational cost. As such we focus on the initial laser-matter interactions and the resulting excited electron dynamics.  For long-term dynamics and equilibration, one could use a final simulation state from PIC as the initial conditions for another model. For example, one could use the electron and ion temperatures in a two temperature model, or consider tabulated equation of state values~\cite{Chimier_PhysRevB.84.094104}.

EPOCH includes multiple physics modules, but was not designed with laser-dielectric interactions in mind. Towards this goal, we have added a new molecular photoionization model and a model for optical material properties as discussed in the next subsections.

\subsection{Keldysh Photoionization}
Our work extends the existing ionization framework already available in EPOCH~\cite{arber2015contemporary} to include the photoionization model developed by L.~V.~Keldysh~\cite{keldysh1965ionization}. For a laser with electric field amplitude $E$ and frequency $\omega$ interacting with a material having band gap $\Delta$ and reduced electron-hole mass $m^*$, the Keldysh parameter is $\gamma = \omega\sqrt{m^*\Delta}/eE$, where $\gamma >> 1$ is the multiphoton ionization regime and $\gamma << 1$ is in the tunneling regime \cite{keldysh1965ionization}. The Keldysh formulation is especially useful as it spans both regimes, which are often present when considering a laser induced damage experiment. For example, the peak electric fields in our simulations give values of $\gamma$ ranging from about 0.3 to 0.9, which means the photoionization regime is neither multiphoton nor tunneling, suggesting that the full Keldysh formula should be utilized.

Now we introduce the ionization rate equation used in our work. For brevity in the following expressions, we follow \cite{Tien_1999_PRL,balling2013femtosecond} by introducing the variables $\gamma_1 = \gamma^2/(1+\gamma^2)$ and $\gamma_2 = 1/(1+\gamma^2)$. The effective band gap is then given by
\begin{equation}
x=\frac{2}{\pi} \frac{\Delta}{\hbar \omega} \frac{\epsilon(\gamma_2)}{\sqrt{\gamma_1}}. 
\end{equation}

We may then write the ionization rate $W$ as
\begin{align}
    W [m^{-3} s^{-1}] &= 2 \frac{2\omega}{9\pi}\left(\frac{m^*\omega}{\hbar\sqrt{\gamma_1}} \right)^{3/2} \\ \nonumber &\times Q(\gamma,x)\exp\left( -\pi \lfloor x+1  \rfloor \frac{\kappa(\gamma_1)-\epsilon(\gamma_1)}{\epsilon(\gamma_2)} \right),
\end{align}

where $\kappa$ and $\epsilon$ are complete elliptic integrals of the first and second kind,  $\Phi$  is the Dawson integral, and 

\begin{align}\label{eq:Q}
Q(\gamma,x) = \sqrt{\frac{\pi}{2\kappa(\gamma_2)}}\sum_{n=0}^\infty\exp\left(-n\pi \frac{\kappa(\gamma_1)-\epsilon(\gamma_1)}{\epsilon(\gamma_2)}\right) \\ \nonumber \times\Phi \left(\sqrt{\frac{\pi^2(\lfloor x+1\rfloor - x + n)}{2 \kappa(\gamma_2)\epsilon(\gamma_2)}} \right).
\end{align}

We note that this is the corrected version of the formulation, where the original contains a misprint as noted by \citet{gruzdev_2004}. Using the uncorrected version can result in significantly different calculations\cite{gruzdev_2004,zhang2022ultrafast}.

For our simulation framework, the Keldysh ionization rate is evaluated in-situ\footnote{with the exception of the elliptic integrals which are read from tabulated data files with 1,000 points} rather than interpolated from a tabulated form which facilitates accuracy over a wide range of fluences. We use the first 500 terms in this infinite sum in Eq.~\ref{eq:Q}, which is sufficient for the regimes of interest in this paper, while limiting the computational cost. An adaptive approach to a prescribed tolerance could be employed in the future. Currently each molecule can only be ionized once.

While the Keldysh parameter depends on the laser amplitude $E$, the amplitudes of our pulses vary within the laser envelope and a given computational cell in a PIC simulation only considers the instantaneous electric field. To address this, we store the electric field magnitude at each time step using an array large enough to store an entire laser cycle on a rolling window, as suggested in \cite{zhang2022ultrafast}.  Then the maximum field for the previous cycle is used to approximate the amplitude to calculate the ionization rate. This approach would encounter challenges for single-cycle pulses. In the future a more complex envelope model could be explored\cite{terzani2019fast}.

\subsection{Collisional Effects}
We use the P\'erez/Nanbu\cite{perez_2012_collision,NANBU1998639} binary collision module already included in the EPOCH code to account for collisions. The collision frequency in EPOCH is calculated for a charged particle $\alpha$ with charge $q_\alpha$ scattering off a charged particle $\beta$ as
\begin{equation}
    \nu_{\alpha \beta} = \frac{(q_\alpha q_\beta)^2 n_\beta \ln (\Lambda)}{4 \pi (\varepsilon_0 \mu)^2} \frac{1}{v_r^3},
\end{equation}
where  $n_\beta$ is the density (for particles of species $\beta$), $\ln (\Lambda)$ is the Coulomb Logarithm, $\mu = m_\alpha m_\beta/(m_\alpha+m_\beta$) is the reduced mass, and $v_r$ is a relative velocity\cite{arber2015contemporary}.  EPOCH extends the model to low temperatures following an approach by \citet{Lee_More_1984,perez_2012_collision}. More details can be found in \citet{arber2015contemporary} and in the source files and documentation provided with the open-source EPOCH code\footnote{\href{https://epochpic.github.io/}{https://epochpic.github.io/}}.  Collisions are included between all charged particles in the simulation. The Coulomb logarithm is calculated automatically with a fixed lower bound of~1.

EPOCH includes a collisional ionization routine designed for atoms\cite{arber2015contemporary,morris_2022_collisional_epoch}, but this is not well suited for impact ionization in dielectrics. A more appropriate ionization rate for our regime can be calculated with the approach by \citet{keldysh1965impact}, although some of the input parameters to the models are not well reported and  may require fitting of output results to experimental data \cite{charpin2024simulation,DRE_Deziel_21}. Impact ionization is reduced for shorter few-cycle pulses. Models by \citet{Petrov_2008} suggest collisional ionization dominates photoionization for fluences exceeding 0.4~J~cm$^{-2}$, whereas the MRE model by \citet{rethfeld2006free} predicts this threshold to be 10~J~cm$^{-2}$\cite{balling2013femtosecond}. For this work, we only consider few cycle (7~fs) pulses and do not include impact ionization. We find good agreement with previous experiments without the need for impact ionization, but expect this to be an important consideration for longer pulses in future work.

\subsection{Refraction}   
The optical properties of the dielectrics are modeled using a spatially varying permittivity $\varepsilon$ throughout the simulation box. At the beginning of the simulation the optical properties are stored in a matrix with the same size as the simulation grid. We then modified the field solver to include a spatially-dependent permittivity when advancing Maxwell's Equations, follow a similar approach to the one in the WarpX code\cite{warpx_fedeli_2022} and the modification to EPOCH by~\citet{charpin2024simulation}. To easily account for arbitrary target structures, we define the shape of the optical region at the same place particle species are initialized. There has been work including nonlinear optical effects in PIC or FDTD simulations\cite{VARIN201870,deziel_2018}, although those effects are not considered here.

\section{Simulation Setup}\label{sec:setup}
\begin{figure}\centering
\includegraphics[width=1\linewidth]{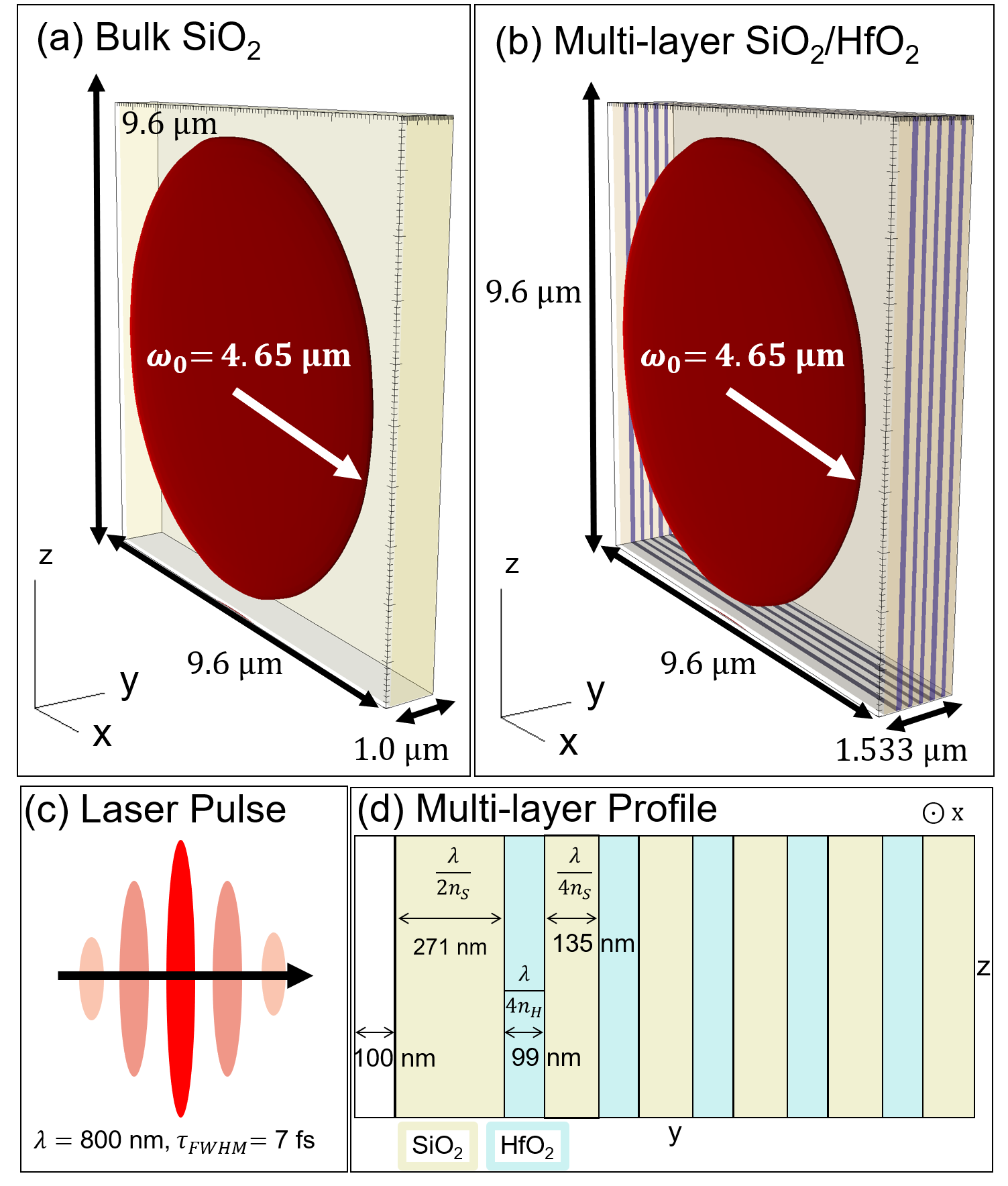}
\caption{A schematic of the 3D simulations for bulk fused silica (a), and the coating of the multi-layer quarter-wave mirror with a fused SiO$_2$ protective layer (b). The normally incident few-cycle laser pulse (c) is driven from the minimum y boundary into a 0.1~$\mu m$ vacuum region before the target. The SiO$_2$ regions are represented in yellow and the HfO$_2$ in blue. A cross section of the multi-layer mirror in the $x-z$ plane is shown in (d) ($z$ dimension is not to scale). The thickness of the top SiO$_2$ layer is 270.90 nm and then there are alternating layers of HfO$_2$ (98.95 nm) and SiO$_2$ (135.45 nm). }\label{fig:Setup}
\end{figure}

We begin by applying this simulation framework to a slab of fused silica corresponding to the experiment discussed in \citet{Chimier_PhysRevB.84.094104} and then apply the framework to a multi-layer mirror as shown in Fig.~\ref{fig:Setup}. We test a range of fluences near the reported damage threshold, where the fluence is given by $F = 2 E_{las}/\pi \omega_0^2$, with $E_{las}$ being the energy of the laser pulse. The laser is introduced as a boundary condition and enters into vacuum before interacting with a slab of fused silica at normal incidence. For both the bulk and multi-layer targets, an $\lambda = 800$~nm, 7~fs Full Width at Half Maximum (FWHM) sine-squared pulse with a spot radius of $\omega_0 = 4.65~\upmu$m is modeled. 

\subsection{Grid and Particle Initialization}
The simulation setup for bulk SiO$_2$ is shown in Fig.~\ref{fig:Setup}(a).  The simulation box is 1.0~$\upmu$m in the longitudinal ($y$) direction and 9.6~$\upmu$m in the transverse ($x$/$z$) directions. The target is 0.9~$\upmu$m thick, leaving 0.1~$\upmu$m vacuum before the laser interacts with the target. Simple outflow boundaries are used to allow transmission of the laser from the boundaries with minimal reflection. The bulk silica simulations have a resolution of 5~nm in the longitudinal ($y$) direction and 40~nm in the transverse directions. A higher resolution was used in the transverse direction to better resolve the ionization dynamics at the interface between the target surface and the vacuum region.

Then we apply this framework to multi-layer interference coatings composed of alternating fused SiO$_2$(yellow) and HfO$_2$(blue) layers as shown in Fig.~\ref{fig:Setup}(b,d). The simulation space is 9.6 $\mu$m by 1.533 $\mu$m by 9.6 $\mu$m long in x, y, z respectively, with resolution of 13.8~nm in the longitudinal ($y$) direction and 40~nm in the transverse directions. The thickness of the surface protective SiO$_2$ layer is $\lambda$/(2n), while the rest of the layer thickness is $\lambda$/(4n), a typical Bragg quarter-wavelength mirror, where $n$ is the index of refraction for each material. The top fused SiO$_2$ layer is placed at 0 $\mu$m and the pulse enters normally from y-axis at –0.1 $\mu$m. The optical path of the source is more than 4 times the length of the longitudinal direction in simulation space.

Both series of simulations are initialized with 1000 neutral SiO$_2$ (or HfO$_2$) macroparticles per cell with a temperature of 300~K. Similar to \citet{charpin2024simulation}, we found a large number of particles per cell were required for accuracy with the Keldysh ionization model in this regime. The simulations are run to a simulation time of 24 fs so that the optical path is about 3 times longer than the y-axis of the simulation box. For each simulation, we use the default time step in EPOCH of 0.95 times the Courant–Friedrichs–Lewy (CFL) limit\cite{CFL_Paper,CFL_Paper_English}, or $0.95/(c\sqrt{1/\Delta x^{2} +1/\Delta y^{2} +1/\Delta z^{2} })$, where $\Delta x,y,z$ represents the grid size in each simulation dimension\cite{arber2015contemporary}.

\subsection{Material Properties}
The simulations require a number of material properties as inputs to model the interaction and interpret the predictions. These include the linear refractive index, molecular number density, band gap, for which we use standard values listed in Table~\ref{tab:mat_prop}. There is less agreement in reported values for the effective electron and hole masses ($m_e^*$ and $m_h^*$ respectively) and subsequently this results in different reduced effective masses $m^* = 1/(1/m_e^* + 1/m_h^*)$. This variation leads to significant differences in predictions for the excited electron density using the Keldysh ionization model~\cite{gruzdev_2010,Zhang_2020}. To calculate $m_e^*$ for fused HfO$_2$, we assume it is the spherically averaged effective mass around the $\Gamma$ and B point of the monoclinic HfO$_2$ ~\cite{Zhang2021,Prevalov2007}. We generally use material properties for m-HfO$_2$ as those for amorphous HfO$_2$ are less readily available. These effective electron masses are used for ionized electron particles in the simulations.

\begin{table*}[htb]
\centering
\caption{Material Properties used in simulations.}\label{tab:mat_prop}
\begin{tabular} {c | c|c|c|c|c}
Material & Band Gap (eV) & Index $n$ & Density (g cm$^{-3}$) & $m_h^*$& $m_e^*$ \\
\hline
SiO$_2$ & 9~\cite{penano2005transmission}  & 1.477 & 2.2~\cite{haynes2012crc}& 8~\cite{gritsenko1995electronic} & 0.6~\cite{gritsenko1995electronic} \\
HfO$_2$ & 5.7~\cite{Prevalov2007} &2.021  & 9.68 & 1.12~\cite{Prevalov2007} &1.09~\cite{Prevalov2007} \\
\end{tabular}
\end{table*}

\section{Damage Modeling of Bulk Silica Target}\label{sec:bulk}

We benchmark our framework against the experiment in \citet{Chimier_PhysRevB.84.094104}\cite{uteza2011control}, which finds damage with a fluence of 1.18~J~cm$^{-2}$ and ablation at 1.3~J~cm$^{-2}$ for a 7~fs FWHM pulse at normal incidence.  There is some uncertainty in these thresholds. Other experiments of LIDT for bulk silica with different experimental conditions including a shorter 5~fs pulses\cite{Lenzner_1998,Kafka:16} report thresholds from 1.5 to 1.8~J~cm$^{-2}$.

\subsection{Electron Density}
\begin{figure}\centering
\includegraphics[width=0.48\textwidth]{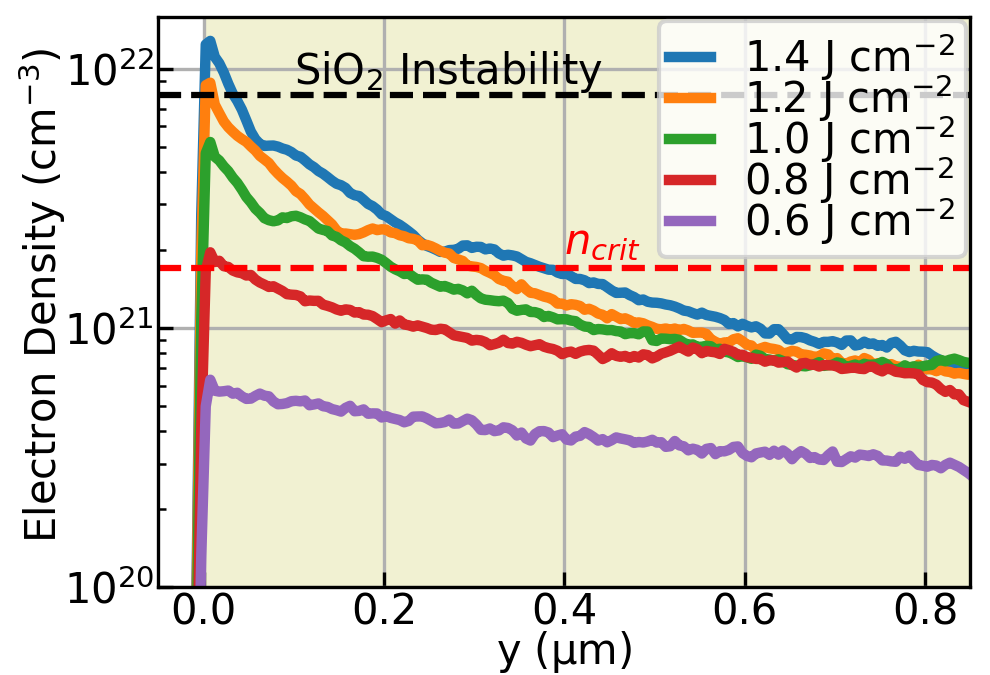}
\caption{The electron density at the center of x-z plane (Averaged over 6 cells in x and  y) along y at 20~fs for a series of PIC simulations at various laser fluences with a 7~fs pulse interacting with bulk SiO$_2$. The critical plasma density $n_{crit}$ and electron instability density from \citet{PhysRevB.42.7163} are labeled with dashed lines. The latter predicts damage around 1.2 J cm$^{-2}$.  }\label{fig:bulk_electron_density}
\end{figure}

For simulation and theoretical work, the predicted excited electron density is often used as a criterion to predict damage. Many studies use the critical electron density\cite{chen2016introduction} for free electrons or some fraction of total ionization \cite{werner2019single}. This qualitatively makes sense as exceeding the s can result in high absorption and subsequent damage. This choice has shown good agreement with longer pulses, although recent work has suggested that this description is insufficient for modeling shorter few-cycle pulses \cite{zhokhov2018optical,Chimier_PhysRevB.84.094104}. As shown in Fig.~\ref{fig:bulk_electron_density}, the damage threshold is predicted at about 0.8~J~cm$^{-2}$ by the critical density criterion, which is about a 30\% underestimation of the experimental LIDT threshold. Number density is a standard output variable for PIC simulations. It is calculated by mapping the position of the macroparticles to the spatial grid based on the shape function selected for the simulation\cite{BirdsallLangdon2004,hockney1981PIC,arber2015contemporary}.

Alternatively, the instability density suggested by \citet{PhysRevB.42.7163} states that when the conduction band electron density reaches about 9\% of valence band electron density, the elastic shear constant will become negative and the lattice becomes unstable, which then leads directly to a very rapid melting of the crystal structure. This criterion was developed for crystals, but the fundamental physical principles—namely, the relationship between conduction band electron density and the stability of the atomic structure is similar. 
By applying this criterion to the bulk fused SiO$_2$, the damage is achieved at about 1.2~J~cm$^{-2}$, which agrees with the experimental results well.

\subsection{Energy Density}

The excited electron energy density criterion has also been suggested to predict damage \cite{grehn2014femtosecond_energy_density}. The predicted energy density is typically compared to material properties such as the dissociation energy, or an energy barrier associated with melting, or boiling\cite{WANG2017_melt_boil,zhokhov2018optical}. We compare the energy densities in our simulations to these thresholds. 

Previous computational approaches typically assume some simple electron energy distribution, whereas in PIC, the energy density can be calculated directly with standard outputs. For our simulations, we multiply the number density by average particle energy for a species in each cell. Alternately, this could be re-sampled to a finer or courser grid if the individual macroparticle positions and energies are extracted from the simulation.

Due to variations of reported material properties in the literature and uncertainty of previous simulations, the exact energy density for damage is not agreed upon. For reference, the dissociation energy of SiO$_2$ has reported values from $54 - 68$~kJ cm$^{-3}$ (Refs.~\cite{jia_2004_energy_density,grehn2014femtosecond_energy_density,Inaba_1999}). For comparison, the threshold for high energy density physics\cite{drake2006introduction} is $\sim$100~kJ~cm$^{-3}$. 

The energy densities related with melting or boiling are lower, where we can use the temperature-dependent heat capacity and latent heat of vaporization used in \citet{zhao_capacity,Zhao_capacity_2019} to calculate an energy density of 5.7 kJ cm$^{-3}$ for melting and 34.7 kJ cm$^{-3}$ for boiling. We do have uncertainty in these values. The latent heat of vaporization has the largest contribution to the boiling criteria and there is a great deal of uncertainty in reported values. For example, the calculation above uses values from \citet{bauerle2013laser} who report a calculated value for the latent heat of vaporization of c-SiO$_2$ to be 1.23$\times 10^7$ J~kg$^{-1}$, while \citet{Kraus_silica_2012} reports $1.177\pm 0.095$ $\times 10^7$ J~kg$^{-1}$, and \citet{khmyrov2014possibility} use 0.96 $\times 10^7$ J~kg$^{-1}$ (from Refs.~\cite{grigoriev1997handbook,samsonov2013oxide}). This gives a range from 28.7 kJ cm$^{-3}$ to 35.6 kJ cm$^{-3}$ for boiling
We expect some uncertainty in melting energy density as well. 

\begin{figure}\centering
\includegraphics[]{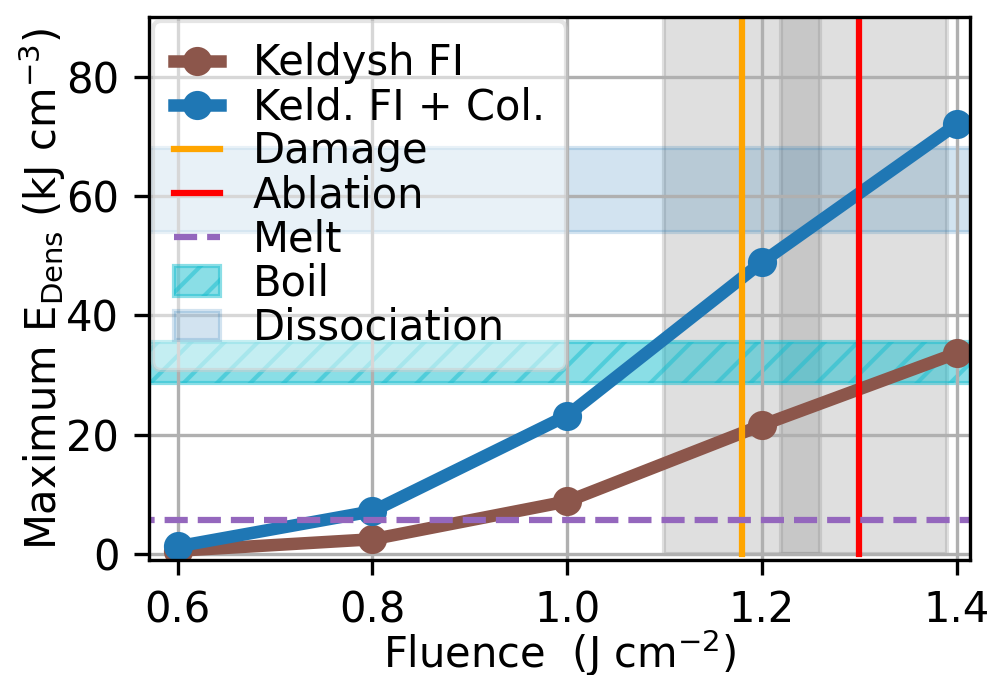}
\caption{The peak energy density for the series of PIC simulations at various fluences with a 7~fs pulse interacting with bulk fused silica. The experiment\cite{Chimier_PhysRevB.84.094104} being modeled observed damage around 1.18~J~cm$^{-2}$ and ablation\cite{uteza2011control} at 1.3~J~cm$^{-2}$. The shaded horizontal line/bands indicate approximate energy density thresholds for melting, boiling, and dissociation. The shaded vertical bands represent uncertainty in experimental damage and ablation thresholds. The simulations including collisions have a higher predicted energy density. Including both our Keldysh photoionization model and collisional effects show agreement between the expected damage fluence and the dissociation energy.   }\label{fig:fluence_ed}
\end{figure}

Figure~\ref{fig:fluence_ed} shows the maximum energy density in simulations with and without collisions for a range of laser fluences.  The simulations just including photoionization, (without collisions) have lower energy density at the end of the simulation and do not exceed the boiling criterion until much higher fluences than are expected for LIDT in these conditions. 

In Fig.~\ref{fig:fluence_ed}, we see that at 1.3~J~cm$^{-2}$, where ablation is observed in experiments, the simulation energy density overlaps with the reported dissociation energy values.  For 1.2~J~cm$^{-2}$ near the LIDT threshold, the energy density for the simulation is around 49~ kJ cm$^{-3}$, exceeding the boiling threshold and near the dissociation energy.

\subsection{Kinetic Particle Motion}
\begin{figure}\centering
\includegraphics[width=0.48\textwidth]{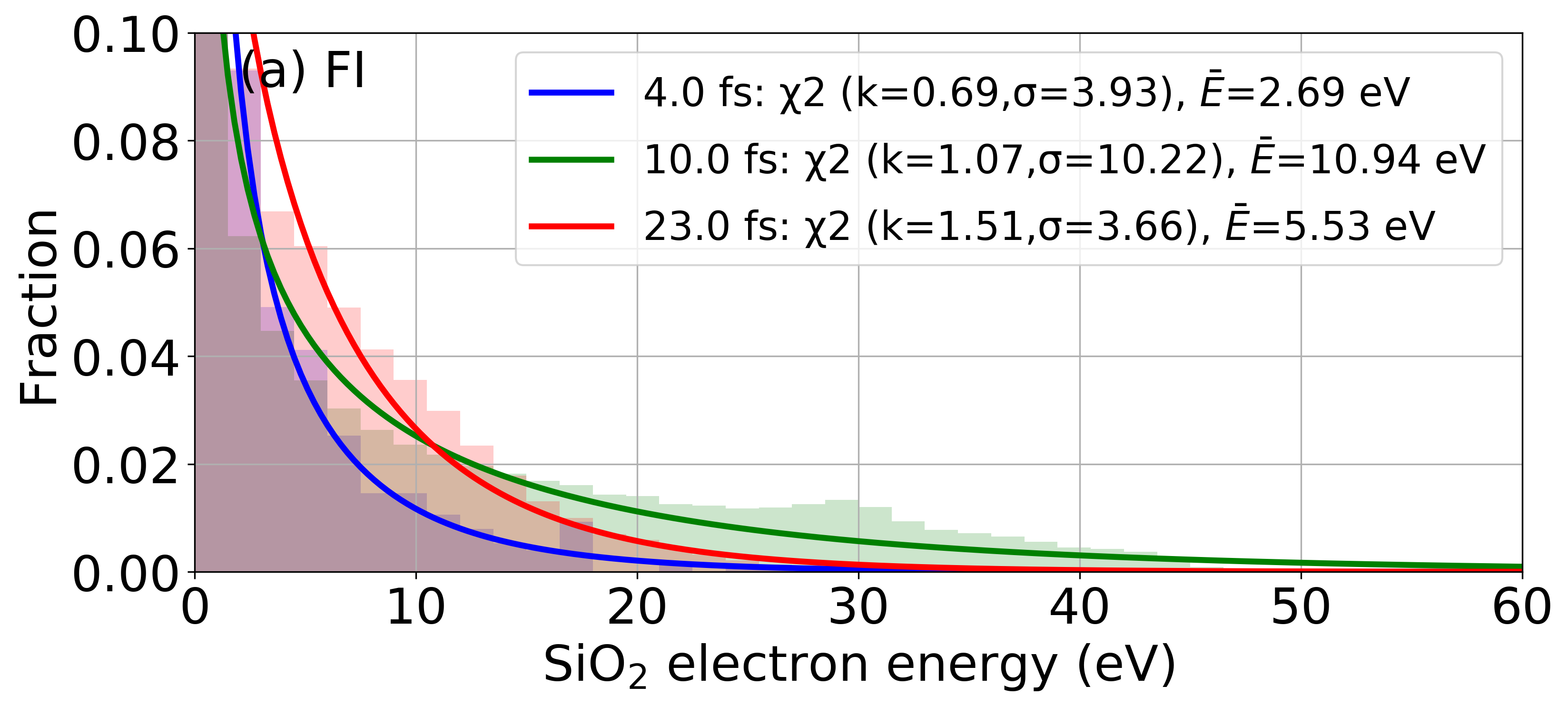}
\includegraphics[width=0.48\textwidth]{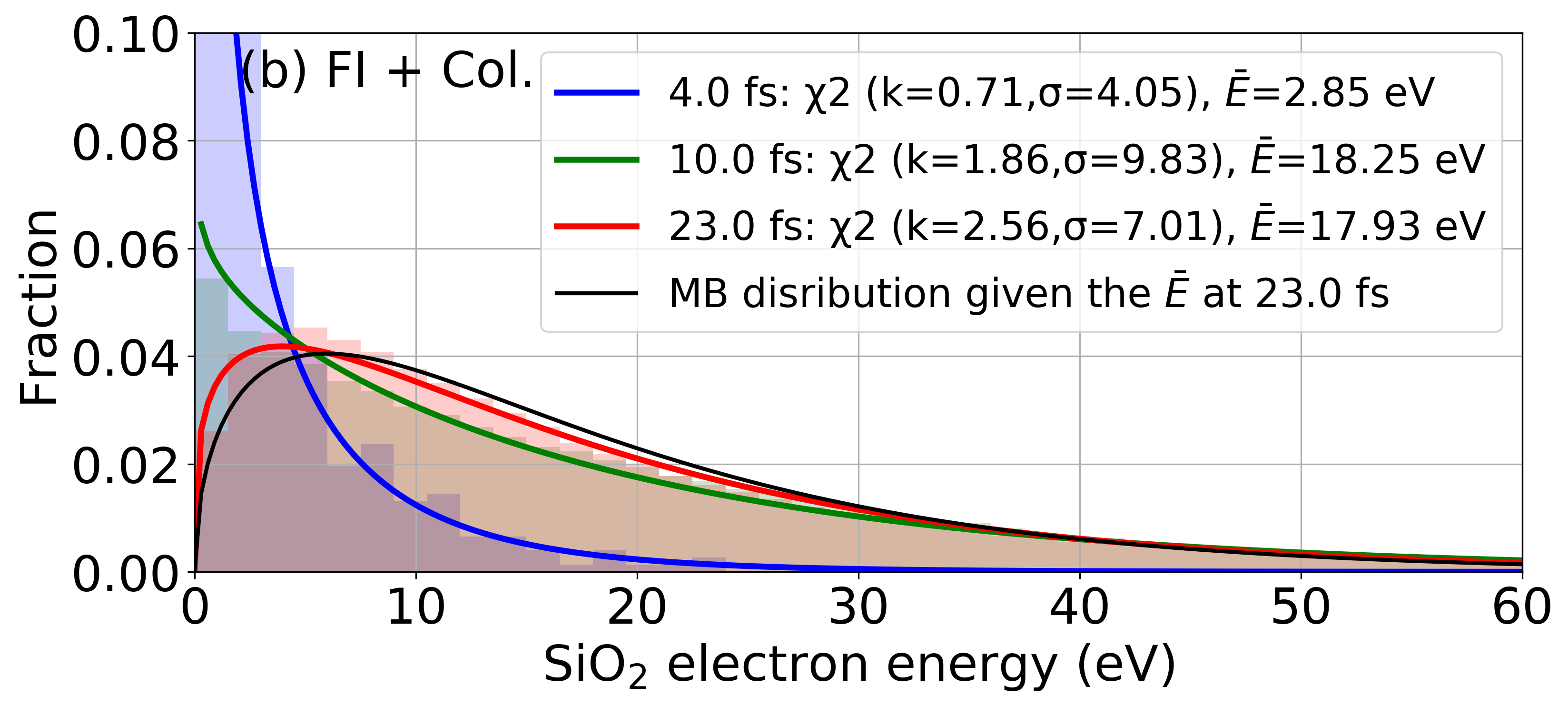}
\caption{Excited electron energy distributions in a (200~nm)$^2\times$ (900~nm) region of the target at the center of the interaction for a 1.2 J cm$^{-2}$ pulse. A simulation with just field ionization is shown in (a) and a simulation with field ionization and collisions is shown in (b). A fit with the given temperature is shown. We observe the nonthermal nature, especially for early times and for simulations without collisions.  }\label{fig:part_dist}
\end{figure}

The kinetic nature of PIC simulations allows us to explore the energy and motion of the excited electrons. While most previous approaches assume a thermal distribution of the excited electrons, Fig.~\ref{fig:part_dist}, shows that this is not the case during the interaction. The spectra at different times of the simulation are shown and the energy is fitted by the $\chi^2$ distribution using the SciPy\cite{2020SciPy-NMeth} library
\begin{equation}
\chi^2: f(E_{e};k,\sigma)=\frac{E_{e}^{k/2-1} e^{-\frac{E_{e}}{2\sigma}}}{(2\sigma)^{k/2}\Gamma(\frac{k}{2})}, 
\end{equation}
where $E_e$ is the energy of the electrons, $k$ is the degree, and $\sigma$ is the scale parameter. The $\chi^2$ distribution becomes the standard Maxwell-Boltzmann distribution when $k$=3 and $\sigma$=$k_b T$/2. In our analysis, we primarily focus on the variation of the degrees of freedom $k$ as it is crucial in describing the main characteristics of the Maxwell-Boltzmann distribution. For both cases, with or without collision, the excited electrons are highly nonthermal at the early stage of 4~fs, the degree $k$ is about 0.7. Then after the peak intensity passed through the target at about 10~fs, the energy spectrum is still nonthermal with k=1.07 if the collision is off as shown in Fig.~\ref{fig:part_dist}(a), while in Fig.~\ref{fig:part_dist}(b) the degree $k$ goes to 1.86. At the stable stage of 23~fs, the electrons still remain the nonthermal nature in Fig.~\ref{fig:part_dist}(a), while in Fig.~\ref{fig:part_dist}(b), the $k$ is about 2.56, which is approaching to the Maxwell-Boltzmann distribution as indicated by the black curve given the average energy at 23~fs. 
Therefore, our simulations not only show the dynamic evolution of the excited electron energy spectrum during the interaction, but also show the importance of including collisions to capture particle dynamics.  When collisions are included in the simulations, the energy absorbed by the electrons increases, leading to a higher average energy (Fig.~\ref{fig:part_dist}) and higher maximum energy density (Fig.~\ref{fig:fluence_ed}) at the end of the simulation.

\section{Damage Modeling of Multi-layer Mirrors}\label{sec:multi-layer}

For the multi-layer dielectric mirror, HfO$_2$ is expected to have a lower damage threshold than SiO$_2$~\cite{Mero_2005_oxide_breakdown,wang2017transition} due to the lower bandgap. Therefore, the damage may be initiated in the first HfO$_2$ layer. For example, \citet{Talisa:20} found the damage threshold for a four-layer SiO$_2$/HfO$_2$ mirror to be half that of a bulk SiO$_2$ target. Due to the high computational cost of 3D simulations and uncertainty in material properties for HfO$_2$, we explore a simple mirror with a relatively small spot size to gain a better qualitative understanding of the interaction. Future validations with experiment should be coupled with more accurate material property measurements and LIDT measurements of bulk HfO$_2$.

\subsection{Electron Density}
\begin{figure}\centering
\includegraphics[width=0.48\textwidth]{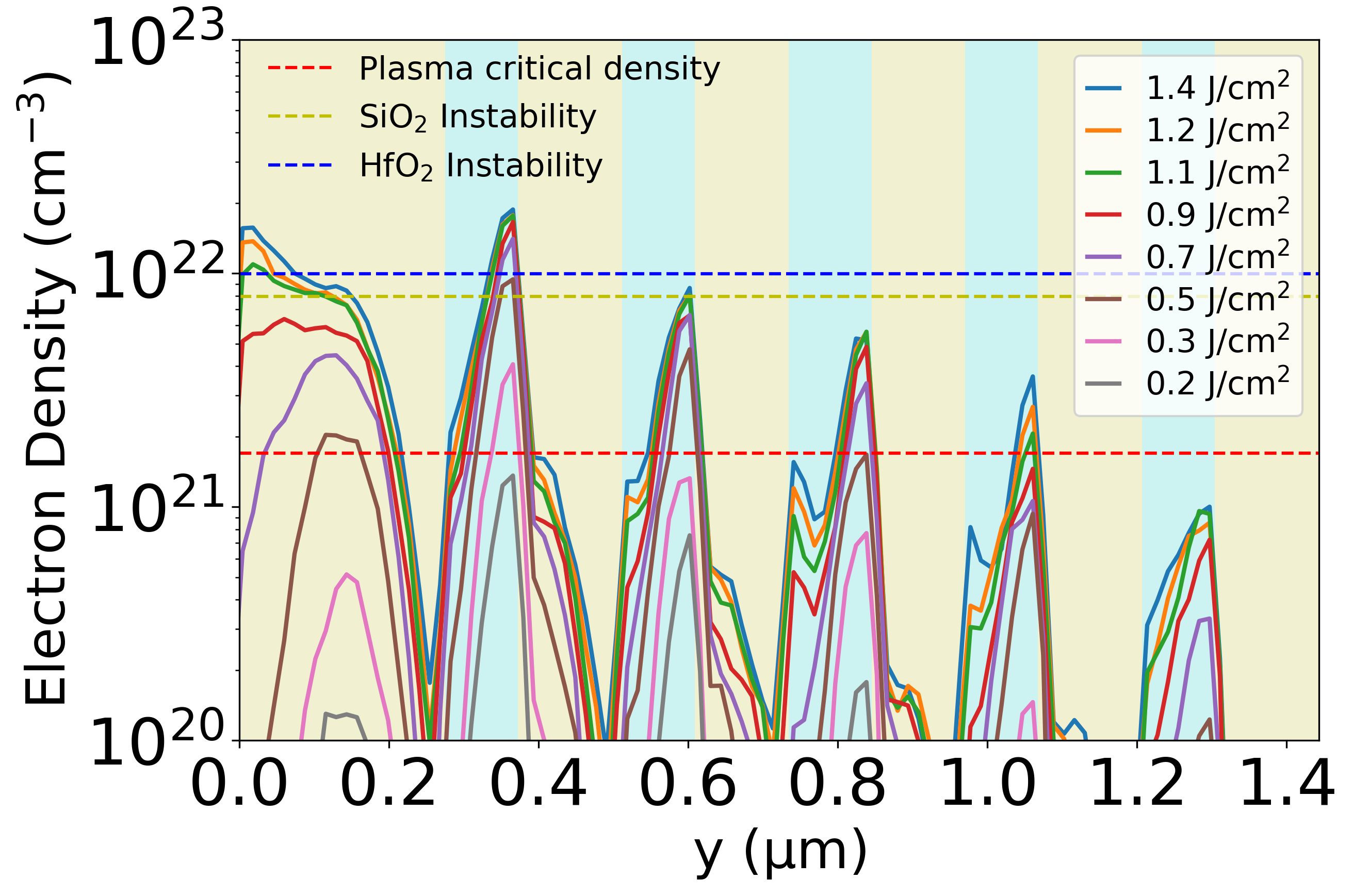}
\caption{The electron density at the center of the x-z plane along y at 20~fs for a series of PIC simulations at various fluences with a 7~fs pulse interacting with multi-layer mirror. The yellow area is SiO$_2$, and blue area is HfO$_2$. Different critical electron densities are labeled in the figure with dashed lines.}\label{fig:mul_electron_density}
\end{figure}

As mentioned in Sec.~\ref{sec:bulk}, the plasma critical density may underestimate the damage threshold, it predicts the LIDT slightly above 0.2 ~J~cm$^{-2}$ in the first HfO$_2$ layer as shown in Fig.~\ref{fig:mul_electron_density}. If we apply the instability density criterion to the mirror target, the damage may be achieved at about 0.5~J~cm$^{-2}$ in the same layer. 
%This prediction is close to the prediction made by the melting energy density criterion. 
Above 0.7~J~cm$^{-2}$, the hot spots in the top two layers are almost fully ionized as the peak electron density maintains at about 1.5~$\times$~10$^{22}$~ cm$^{-3}$.

\subsection{Energy Density}
\begin{figure}\centering
\includegraphics[width=0.48\textwidth]{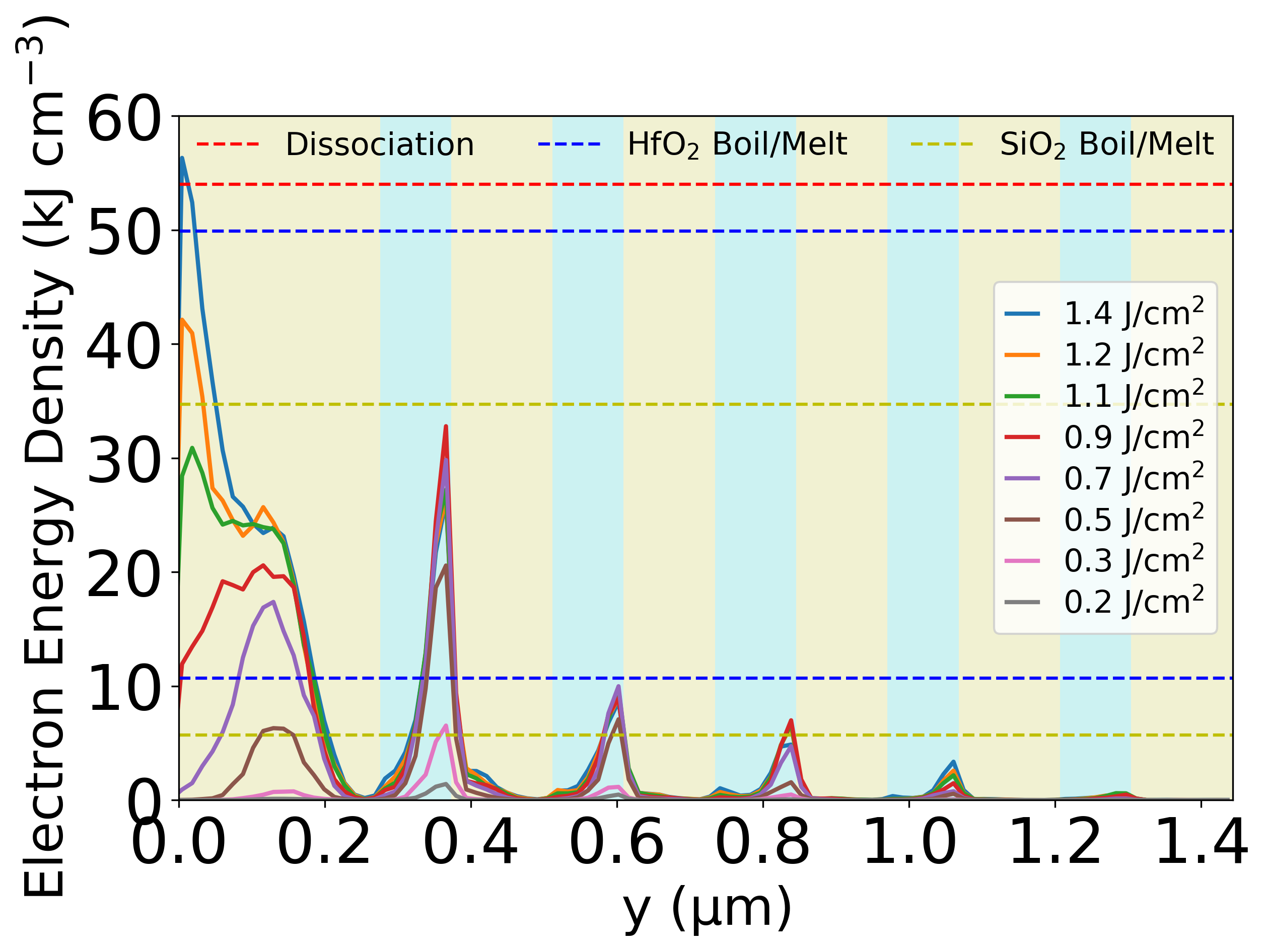}
\caption{The peak electron energy density at the center of x-z plane along y at 20~fs for a series of PIC simulations at various fluences with a 7~fs pulse interacting with multi-layer mirror. The yellow area is fused SiO$_2$, and blue area is fused HfO$_2$. The SiO$_2$ boiling and melting energy density are at about 34.7 kJ~cm$^{-3}$ and 5.7 kJ~cm$^{-3}$ indicated as the yellow dashed lines, and the HfO$_2$ boiling and melting energy density are at about 49.9 kJ~cm$^{-3}$ and 10.7 kJ~cm$^{-3}$ indicated as the blue dashed lines. }\label{fig:mul_energy_density}
\end{figure}

The peak electron energy density along y at the center of x-z plane in each layer with different fluences is shown in Fig.~\ref{fig:mul_energy_density}. The dissociation energy of fused HfO$_2$ is less reported, especially for amorphous samples as we expect in these coatings. There are reports on the formation energy density for m-HfO$_2$ from 48.44 to 52.64~kJ~cm$^{-3}$ \cite{Wang2006,RothBecker1932,Humphrey1953,Huber1968,Kornilov1975,Paputskii1974,Panish_HfO2_1963}, which are similar to the dissociation energy density of SiO$_2$.   
We assume it has the same value with fused SiO$_2$ since both of their molecules have four valence band electrons and are amorphous \cite{zhang2022ultrafast}, which is about 54 kJ~cm$^{-3}$ indicated by the red dashed line. In addition, the effects of structure on boiling and melting points are not considered, as each layer is assumed to retain the melting and boiling points characteristic of its bulk state. This criterion suggests the damage should occur at the surface at fluence of about 1.4~J~cm$^{-2}$, which is even higher than the LIDT of bulk fused SiO$_2$ discussed in Sec.~\ref{sec:bulk}.

Instead let us consider energy density thresholds for melting and boiling as these may relate to damage within the layers of a coating. To calculate the melting and boiling energy densities, we make the following assumptions: (a) the vaporization latent heat for HfO$_2$ is the same as SiO$_2$, and (b) the heat capacity for fused HfO$_2$ is the same as m-HfO$_2$ \cite{osti_1762259}. These approximations give the boiling energy density of HfO$_2$ to be about 49.9 kJ~cm$^{-3}$, which is much higher than that of SiO$_2$ 34.7 kJ~cm$^{-3}$, and may overestimate the actual damage threshold. Similarly, the melting energy density of SiO$_2$ is about 5.7 kJ~cm$^{-3}$, and we calculated that for HfO$_2$ which is about 10.7 kJ~cm$^{-3}$. 

The boiling energy density criterion predicts LIDT at a fluence between 1.1~J~cm$^{-2}$ and 1.2~J~cm$^{-2}$ on the surface of the mirror, which is close to the LIDT of bulk SiO$_2$. Applying melting energy density criterion, the LIDT is predicted to be less than 0.5~J~cm$^{-2}$, and the damage site is initiated in the first HfO$_2$ layer as expected.

\subsection{Plasma Screening Effects}
As shown in Fig.~\ref{fig:mul_energy_density}, the global maximum energy density at low fluences is in the first HfO$_2$ layer as expected. When the fluence exceeds about 1.1~J~cm$^{-2}$ it shifts to the top SiO$_2$ layer and the HfO$_2$ energy density increases slowly after the  fluence reaches at about 0.7~J~cm$^{-2}$. This is because the excited electron density in the first SiO$_2$ layer begins to exceed the critical plasma density (Fig.~\ref{fig:mul_electron_density}). In the top SiO$_2$ layer, the steady state of the dynamic simulation leads to a local maximum enhancement of the electric field at the center. 

As the fluence increases beyond 0.5~J~cm$^{-2}$, we observe a new peak appearing and shifting from the center to the surface in both electron and energy density profile. The ionization rate at the center of the top SiO$_2$ is enhanced due to the strong intensity and thus the electron density reaches the maximum, which further leads to the increased photoionization at this location. The absorption and reflection will continue to increase so that the source can hardly penetrate the target. Therefore, the resonant pattern of the electric field is altered, and a new intensity peak appears.

\begin{figure}\centering
\includegraphics[width=0.49\textwidth]{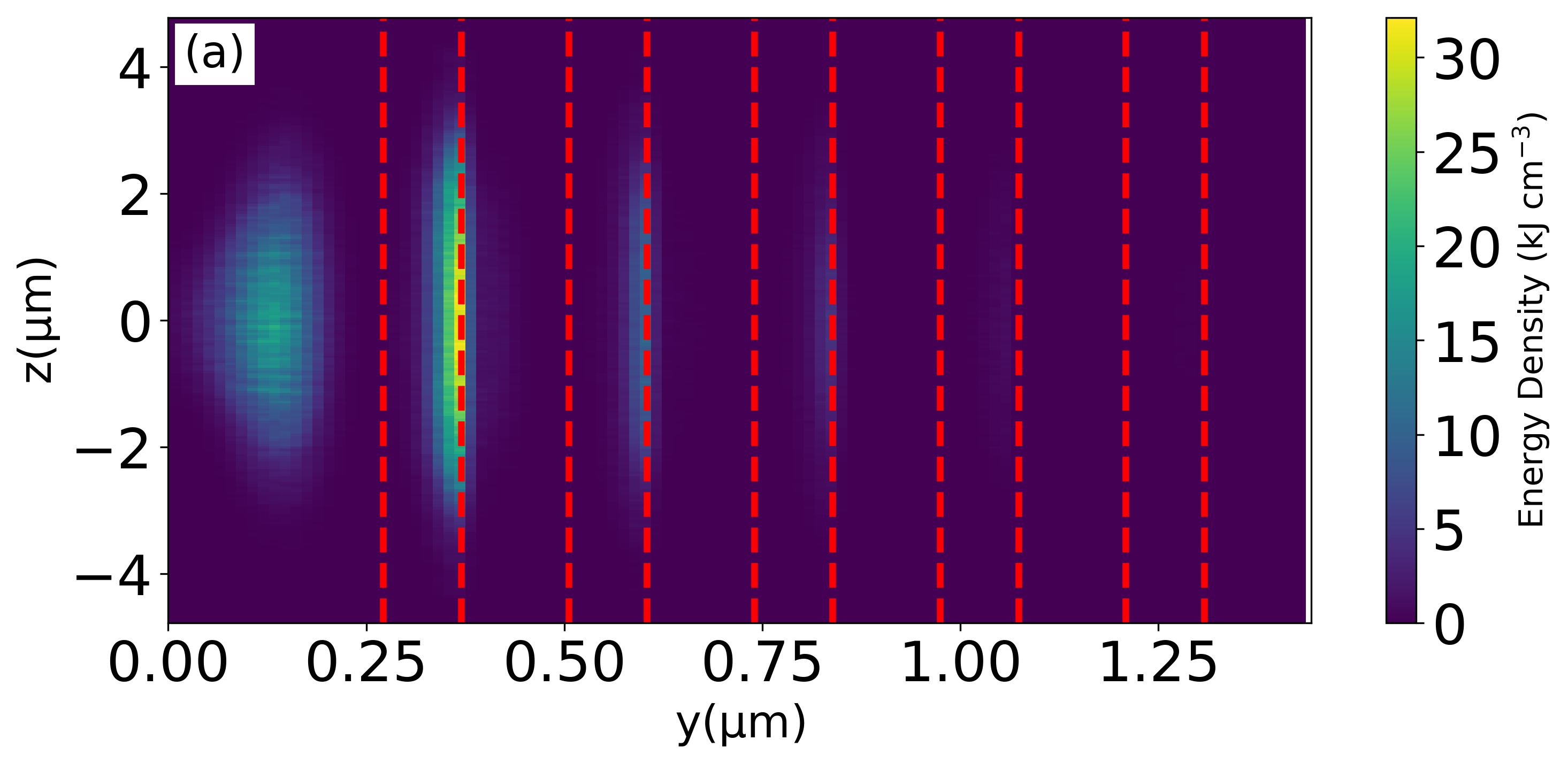}
\includegraphics[width=0.49\textwidth]{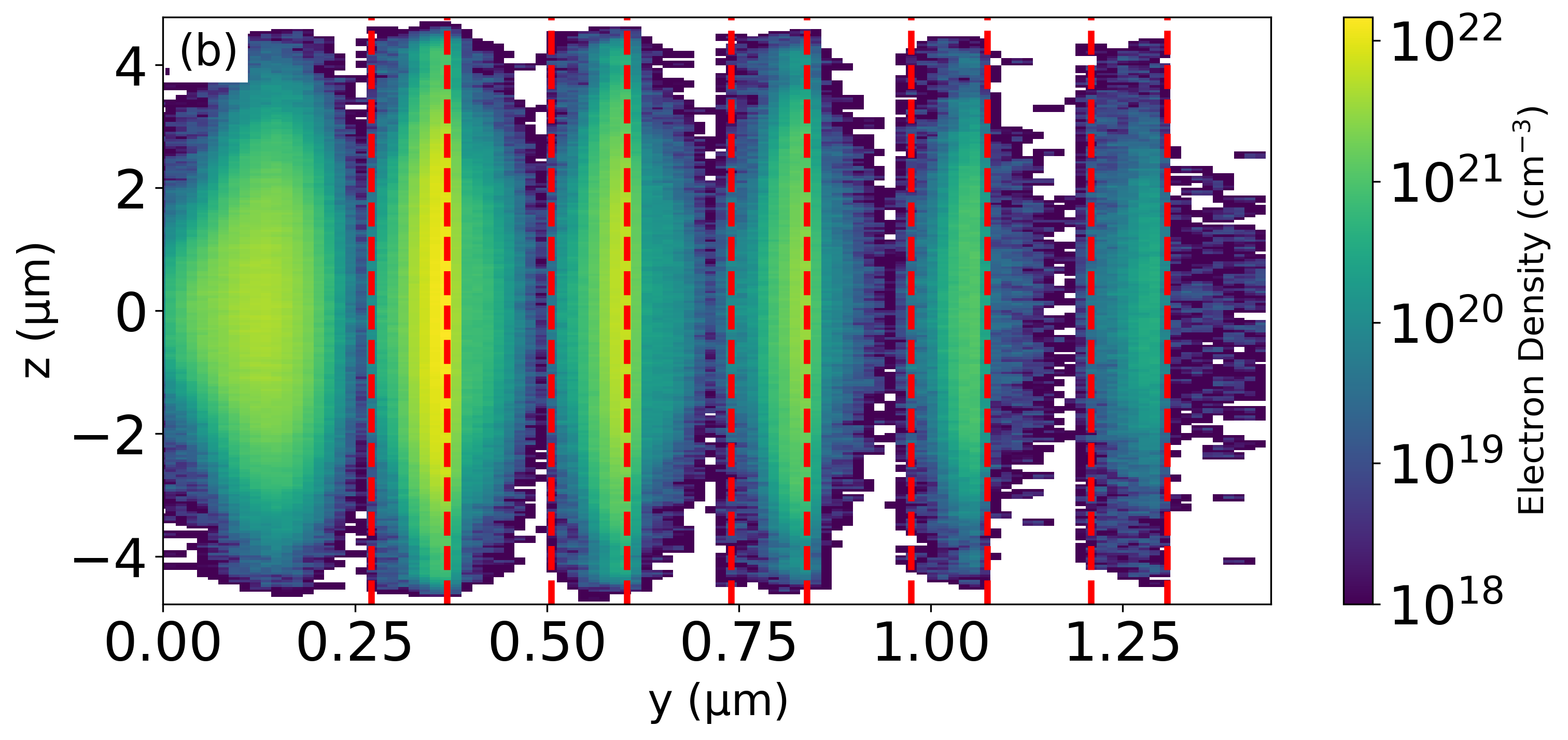}
\includegraphics[width=0.49\textwidth]{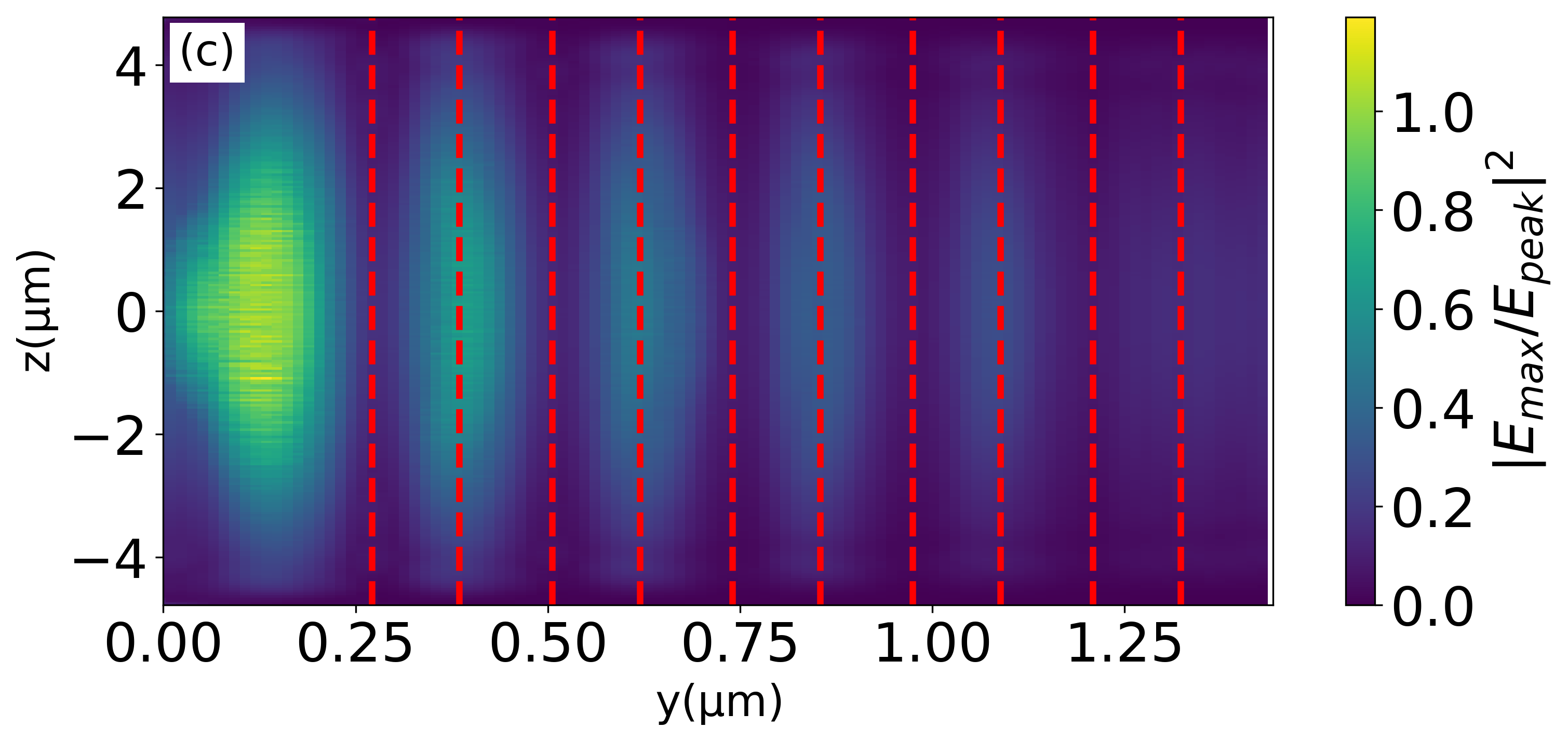}
\caption{The (a) energy density and (b) electron density at the center of x on y-z plane at 20~fs at 0.7~J~cm$^{-2}$ with a 7~fs pulse interacting with multi-layer mirror. The mirror surface starts at y=0, and the dashed red lines are the interfaces between layers. The white areas have no excited electrons. (c) The maximum accumulated normalized intensity over the entire simulation.}\label{fig:energy&density}
\end{figure}

To illustrate this process clearly, a two-dimensional y-z cross section of the layers at the center of x-axis are shown in Fig.~\ref{fig:energy&density} at fluence of 0.7~J~cm$^{-2}$. The maximum electron and energy density are both observed in the first HfO$_2$ layer as shown in (a) and (b), while the maximum accumulated intensity is in the first SiO$_2$ layer in (c). The intensity is recorded at intervals of 0.5~fs and compared across all time points to generate this figure. These results are similar to the results in the FDTD simulation works by \citet{zhang2022ultrafast}.

Noticeably, the electron, energy density, and intensity profiles in the protective SiO$_2$ layer all present a bump extending from the center to the surface, which is different from what previous FDTD simulation works by
\citet{zhang2022ultrafast}. We observed that the electron density peak first appeared at the center and then expanded at about 9~fs to the surface, and became stable at about 15~fs. These are the time when the pulse peak intensity reached the surface and when the entire pulse left the surface. 

Furthermore, in Fig.~\ref{fig:mul_energy_density}, the first HfO$_2$ layer is more strongly affected by the lower fluence pulse. At higher fluence, there is significant growth of the energy density at the surface, while the lower layers do not show as significant of an increase. 
In addition, the energy density at 0.9~J~cm$^{-2}$ is highest among all the fluences, which adds another evidence that the first SiO$_2$ layer has reflected more injected energy due to the plasma screening effects.

The plasma generation could also imply the breakdown threshold which is defined as the permanent change of the optical property. We can see from both energy and electron density profile, the local peak in the top SiO$_2$ layer starts to shift at 0.5~J~cm$^{-2}$ and the profile is no longer symmetric at the center. This analysis is also consistent with the conclusion predicted by the instability criterion.

\subsection{Particle Energy}
\begin{figure}\centering
\includegraphics[width=0.48\textwidth]{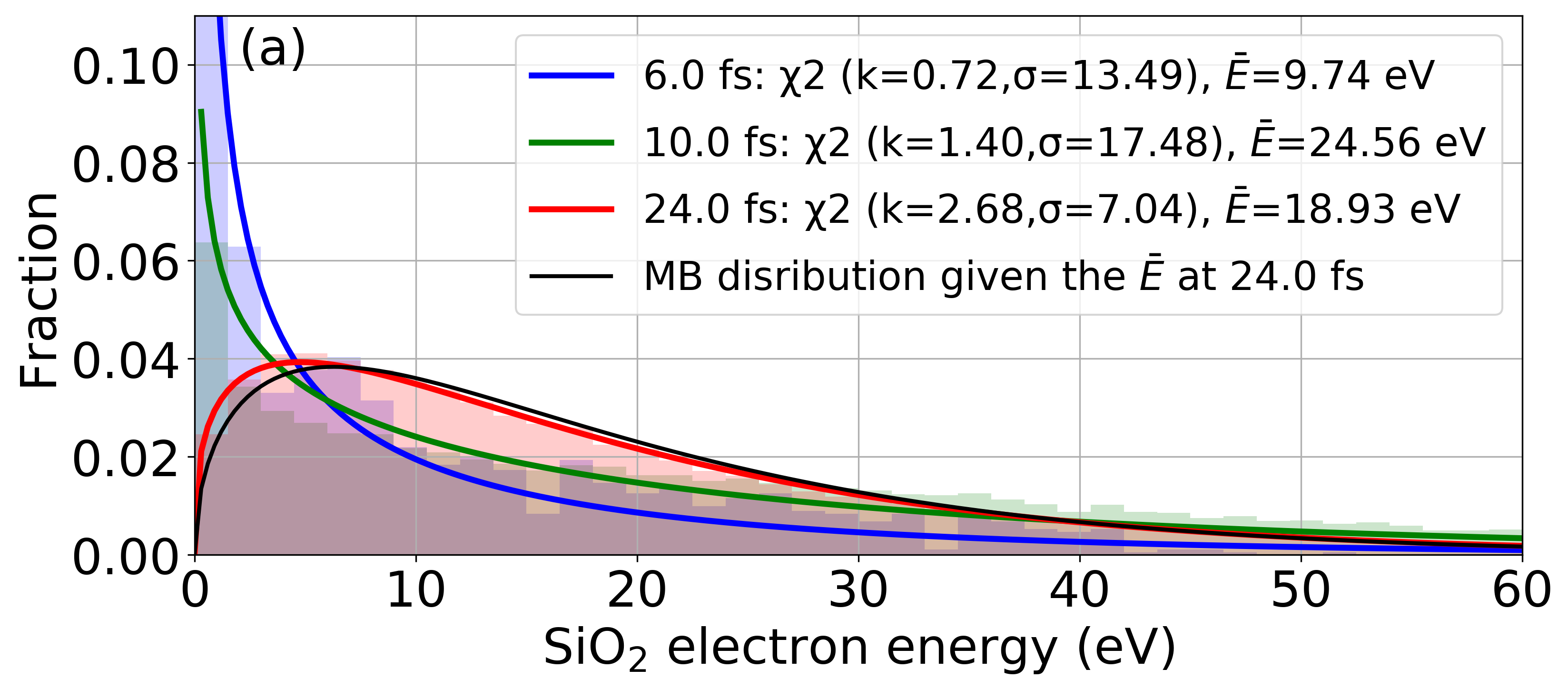}
\includegraphics[width=0.48\textwidth]{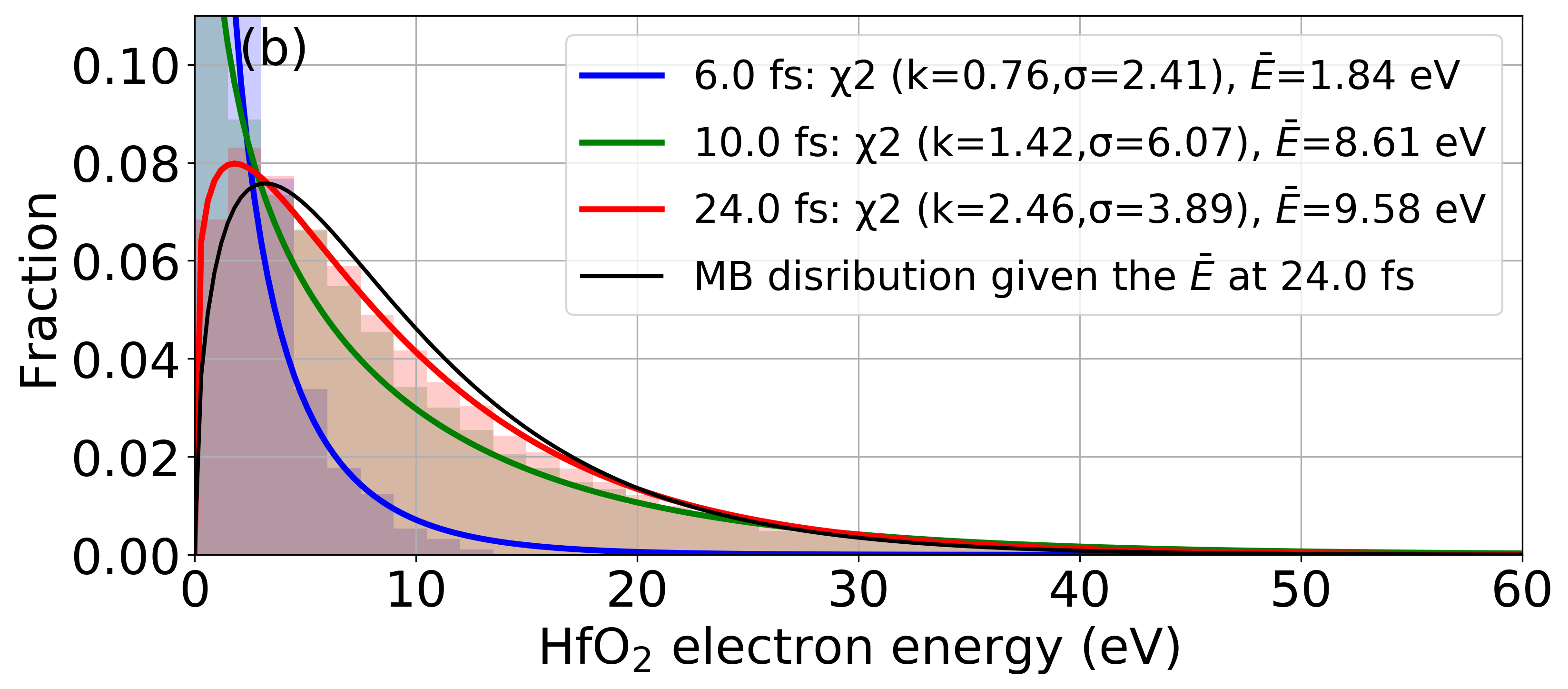}
\caption{The energy histogram for the electrons in the first SiO$_2$ layer (a) and HfO$_2$ layer (b) at 6, 10, 24~fs for the 0.7~J~cm$^{-2}$ simulation  with $\chi^2$ distribution fitted. The black curve is the Maxwell-Boltzmann distribution given the average energy at the stable stage around 24~fs.}\label{fig:hist_particelEnergy}
\end{figure}

The kinetic nature of the excited electrons is shown in Fig.~\ref{fig:hist_particelEnergy}. The spectrum of the electrons in the SiO$_2$ layer is wider than that in the HfO$_2$ layer, though both maximum electron density and energy density reaches maximum in the first HfO$_2$ layer. For both layers, three snapshots of the energy distribution are shown. At the early stage about 6~fs, after a few femtoseconds of interaction, there are some excited electrons generated from the ionization. The electrons are highly nonthermalized since the degree $k$ is less than one. At about 10~fs, the peak intensity has passed through the target, leading to an average electron energy of 24.56~eV and 8.61~eV for the SiO$_2$ and HfO$_2$ layers. The electrons are still nonthermalized as $k$ is about 1.4. At the later stage about 24~fs, the pulse front has left the target for about 10~fs. The degree $k$ is above 2 and approaching to 3, indicating the thermalization process is finishing towards a Maxwell-Boltzmann distribution indicated by the black curve in Fig.~\ref{fig:hist_particelEnergy}.

\subsection{Particle Dynamics}

\begin{figure*}\centering
\includegraphics[width=0.55\textwidth]{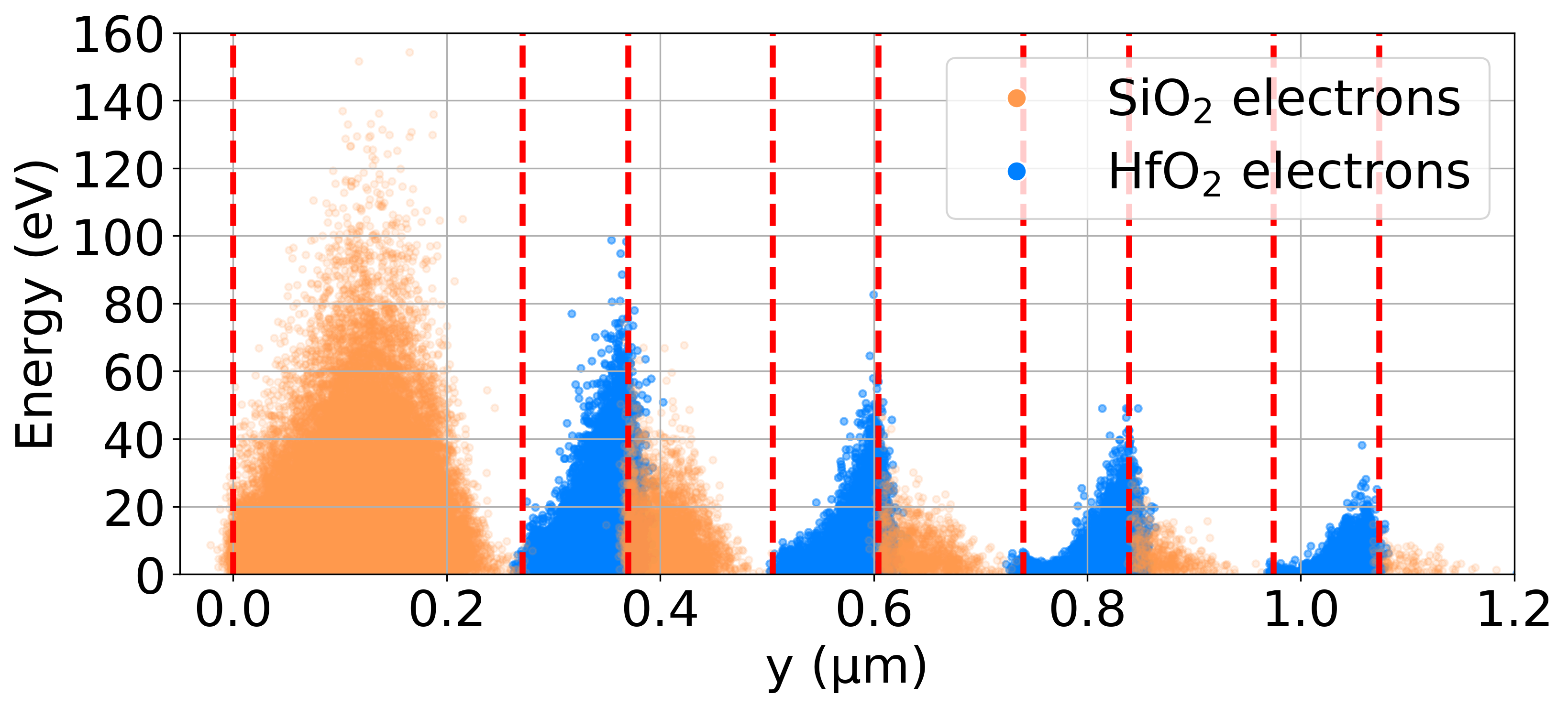}
\includegraphics[width=0.4\textwidth]{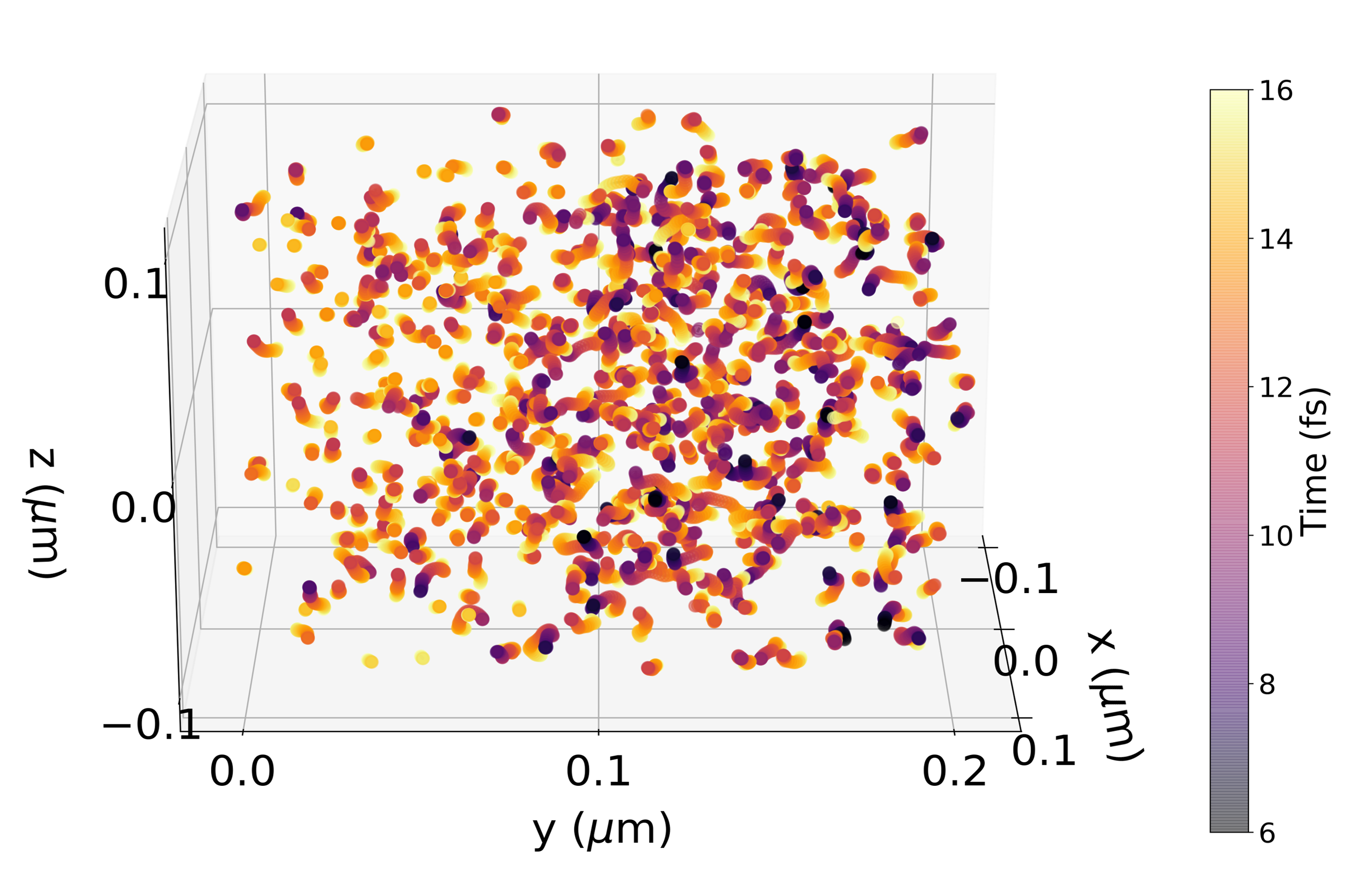}
\caption{The SiO$_2$ and HfO$_2$ electron energy spatial distribution for a fluence of 0.7~J~cm$^{-2}$ is shown on the left at 24~fs. The highest energy particles are found in the center of the first layer and at the interfaces between layers. Particle trajectories for a random 2\% of electrons with low (below average) final energy demonstrate the ionization dynamics in the first SiO$_2$ layer up to 16~fs (after the pulse has left). Ionization near the surface generally occurs at later times as indicated by the color of the tracks. }\label{fig:part_dynamics}
\end{figure*}

The peak intensity of the pulse is about 10$^{14}~W cm^{-2}$, giving a theoretical ponderomotive energy U$_p$ of approximately 10~eV and 5.5~eV for SiO$_2$ and HfO$_2$ electrons respectively (using the $m_e^*$ values from Table~\ref{tab:mat_prop}). In gases, electrons with kinetic energies exceeding  10U$_p$ are typically observed \cite{Paulus1994}, and in our simulation, the highest electron energy from the first SiO$_2$ layer reaches up to 150 eV. Additionally, a few electrons in the first HfO$_2$  layer achieve energies around 100 eV, as shown in Fig.~\ref{fig:part_dynamics} (left). These results align with the kinetic energy predictions for excited electrons based on the Drude model assuming collisional frequency to be 1~fs$^{-1}$ and the energy would be about 118~eV and 65~eV for SiO$_2$ and HfO$_2$, as seen in the studies by \citet{PhysRevB.83.075114} 
Furthermore, the excited SiO$_2$ electrons at the HfO$_2$-SiO$_2$ interfaces rarely get into the HfO$_2$ layers before them, while there are a limited number of HfO$_2$ electrons penetrating into the adjacent SiO$_2$ layers for a few of tens nanometers.   

The tracks of select electrons in the first SiO$_2$ layer are shown from 6 to 16 fs in Fig.~\ref{fig:part_dynamics} (right). Most electrons near the vacuum interface are born after 10 fs, indicated by their yellowish tails. In contrast, electrons near the center of the first SiO$_2$ layer are born earlier. This suggests that the electron density expansion in Fig. \ref{fig:mul_electron_density} and Fig.~\ref{fig:energy&density}(b) is due to direct excitation near the surface rather than displacement from the center.

\section{Conclusion}\label{sec:conclusion}

Understanding few-cycle pulse interactions with dielectric optical components and their LIDT is essential for advancing next-generation laser systems. The use of kinetic simulations in relevant regimes are becoming increasingly popular\cite{Mitchell:15, deziel_2018,Ding_plasmon_2020,charpin2024simulation}. Kinetic simulations allow us to capture the nonthermal nature of the initial interaction, which is important to accurately model absorption and ionization. We show that both excited electron density and energy density provide insight into LIDT. Our framework shows good agreement with experimental LIDT thresholds for bulk silica targets. Multi-layer mirror simulations indicate that plasma screening effects can alter the laser interaction and electron energy distribution for high fluences.

In the future, this framework can be applied to a variety of mirror and grating designs\cite{zhang2022ultrafast} for both near and mid-infrared wavelengths\cite{Austin:18,shcherbakov2023nanoscale}. Additionally, simulations can be inserted into an optimization algorithm to optimize LIDT or other properties of interest for optical components\cite{smith2020optimizing}. This framework already provides a deeper qualitative understanding of the dynamics of laser damage, and we show promising quantitative agreement with experiment for few-cycle pulses.  Further development of this framework including impact ionization\cite{morris_2022_collisional_epoch,charpin2024simulation}, coupled with more precise measurements of material properties for relevant optical coating designs can allow further validation of the framework across large ranges of laser fluence.

\section*{Acknowledgements}
This research was funded by DOE STTR grant no. DE-SC0019900. This research used resources of the National Energy Research Scientific Computing Center (NERSC), a Department of Energy Office of Science User Facility using NERSC award SBIR-ERCAP0021422. This project also utilized resources from the Ohio Supercomputer Center~\cite{OhioSupercomputerCenter1987}. The code EPOCH used in this work was in part funded by the UK EPSRC grants EP/G054950/1, EP/G056803/1, EP/G055165/1 and EP/M022463/1. This work used matplotlib~\cite{Hunter:2007}, SciPy\cite{2020SciPy-NMeth}, and NumPy\cite{harris2020array} to graph and analyze the simulation results.

\bibliographystyle{unsrtnat}
\bibliography{main.bib}

\end{document}